\documentclass[12pt]{article}

 \pdfoutput=1 

\usepackage{amsmath}
\usepackage{amsfonts}
\usepackage{amssymb}
\usepackage{breqn}
\usepackage{subfigure}
\usepackage{color}
\usepackage{cleveref}
\usepackage{amssymb,amsmath,bm,bbm}

\usepackage[margin=1.0in]{geometry}

\begin{document}

\begin{center}
{\bf \Large {A Perspective on Plasticity, Dissipation and the
  $\mathbf{2}^{\rm{\bf {nd}}}$ Law   of
    Thermodynamics}}\footnote{Preprint of an invited perspective paper for the ASME
Journal of Applied Mechanics}  \\ 
\vspace{3mm}
Alan Needleman \\
Department of Materials Science \& Engineering \\
Texas A\&M University,  College Station, TX 77843 USA \\
%\today
\end{center}

\noindent {\bf Abstract} \\
The  requirement of a non-negative dissipation rate for all possible
deformation histories is generally imposed on plastic constitutive
relations. This is  a constraint analogous to the Coleman-Noll
\cite{CN64} postulate that the Clausius-Duhem inequality needs to be
satisfied for 
all possible deformation histories. The physical basis for the Clausius-Duhem
inequality is as a statistical limit for a large number of discrete
events for a long time and is not a fundamental physical
requirement for small systems for a  short time. The relation between the
requirement of a non-negative dissipation rate  and the Clausius-Duhem
inequality  is considered. The consequences of  imposing
a non-negative dissipation rate for all possible deformation histories
are illustrated for: (i) a single
crystal plasticity framework that accounts for elastic lattice curvature
changes as well as elastic lattice straining; and (ii) for discrete defect
theories of  plasticity, with attention specifically on
discrete dislocation plasticity for 
crystalline solids and discrete shear transformation
zone (STZ) plasticity for amorphous solids.  Possible less restrictive 
conditions on  the evolution of dissipation in plasticity
formulations are considered as are implications for stability. The
focus is on
open questions and issues. 

\section{Introduction}

Theories of plasticity, as opposed to theories of thermoplasticity, are based
on conservation laws (momentum, mass and energy) that only involve
deformation 
measures, stress measures and  internal variables. Temperature
only enters through  the 
possible dependence of material parameters on a temperature that is
related to the evolution of plastic dissipation. 

The requirement of a non-negative dissipation rate is a
thermodynamic-like 
requirement analogous to the 2${\rm nd}$ law of thermodynamics as
expressed by the Clausius-Duhem inequality. 
However, the  Clausius-Duhem inequality differs from the basic
conservation  laws in that it  emerges as  a  statistical limit for a 
large number of discrete events over a long time and has a probability
less than one of 
being satisfied when  the  the process under 
consideration involves  discrete events that take place over a 
 sufficiently small  region  for a  sufficiently  short time, see
 e.g. Evans and Searle \cite{Evans02}, 
 Jarzynski \cite{Jarz10}.  
Most investigations where a violation of the 
 Clausius-Duhem inequality has occurred have been at the
 atomistic scale, e.g. Evans et al. \cite{Evans93}, Wang et al.
 \cite{  Wang10}. However,  
 Ostoja-Starzewski and Laudani \cite{Ostoja20} observed a 
violation  of the  Clausius-Duhem inequality in 
 macro scale discrete particle simulations of a  granular
 solid.

Despite these observations, for  continuum formulations at any scale,  the 
Coleman-Noll \cite{CN64} postulate that requires  the Clausius-Duhem inequality 
to be  satisfied  for all possible
deformation histories is typically imposed. At least in
some contexts, 
satisfying this postulate provides a stability condition,
Coleman and Mizel \cite{Cole67}, Dafermos \cite{Dafer79}.  In
developing plastic 
constitutive relations, analogous to the Coleman-Noll  \cite{CN64}
postulate, the plastic dissipation rate is generally required to be non-negative
for all possible deformation histories. 

At the micron scale, plastic deformation almost always, if not
always, is a consequence of the evolution of processes that are
discrete in both space 
and time, such as discrete dislocations  for crystal
plasticity and discrete  shear transformation zones (STZs) for
plasticity of amorphous solids. Imposing the requirement of a
non-negative dissipation rate for the evolution of 
individual discrete defects can lead to  strong 
restrictions, that are not consistent
with atomistic modeling and/or experiment, see e.g. Needleman
\cite{EMC23}.  

These observations suggest that imposing the requirement of a
non-negative 
dissipation rate  for all possible deformation histories on continuum
formulations of plasticity may, in general,  be overly
restrictive. To provide a perspective on this possibility, the following
questions are considered:

\begin{description}

\item{1.} To what extent does the requirement of a non-negative
  dissipation rate emerge from the 2${\rm nd}$ law of thermodynamics? 

\item{2.} What are the consequences of imposing the requirement of a
  non-negative  dissipation rate on a phenomenological theory of
  crystal plasticity?

\item{3.} What are the consequences of imposing the requirement of a
  non-negative  dissipation rate on  discrete defect plasticity
  theories?

\item{4.} For both phenomenological and discrete defect theories of
  plasticity, what are the consequences of imposing the stronger
  requirement  that the plastic dissipation rate is non-negative for all possible
deformation histories?

\item{5.} What are the consequences for stability when the dissipation rate is negative?

\end{description}

To begin, statements of the first and second laws of
thermodynamics are presented.

\section{The $\mathbf{1}^{\rm{\bf {st}}}$  law}

Attention is confined to small deformation kinematics with quasi-static
deformation histories with the only
loading being imposed surface tractions and/or imposed surface
displacements.  Energy densities are written 
per unit reference volume rather than per unit mass. The focus is on a
purely mechanical formulation, i.e. the  independent field variables
entering the 
formulation are stress and displacement, along with their time
and spatial derivatives (such as strain and strain rate), and
possible constitutive internal variables. 

The conservation laws of continuum mechanics -  conservation 
of mass, conservation of linear momentum; conservation of angular
momentum, and conservation of energy - hold for all points of the
body and for all time. In particular, for a purely
mechanical formulation and quasi-static deformation histories,
conservation of energy (the $1^{\rm{\bf {st}}}$  law) for the
body requires  
\begin{equation}
\dot{W} = \dot{\Phi} +\dot{\cal{D}} 
\label{eq0}
\end{equation}
where $W$ is the mechanical work input, $\Phi$ is the energy
stored in the body, ${\cal{D}}$ is the material dissipation in  the
body 
and $(\dot \ )= \partial (\ )/\partial t$. Eq.~(\ref{eq0}) is presumed
to hold for all times.

The quantities  $W$, $\Phi$ and
$\cal{D}$ can be calculated for a body and for any sub-volume of that
body  in terms of volume densities so that
\begin{equation}
W=\int_V w dV \quad , \quad \Phi=\int_V \phi dV \  \  , \  \ 
{\cal{D}}= \int_V \xi dV
\label{eq0ab}
\end{equation}
Here, $V$ is the volume of the body or of the sub-volume.

If Eq.~(\ref{eq0}) holds for any sub-volume of the body
\begin{equation}
\dot{w}= \dot{\phi}+\dot{\xi}
\label{eq11az}
\end{equation}
for all points $x_k$ of the body and for all times.

\section{The $\mathbf{2}^{\rm{\bf {nd}}}$ law}

Consideration of the $2^{\rm nd}$ law requires the introduction of 
entropy and absolute temperature which are assumed to be well
defined even 
though plastic deformation processes are not necessarily thermodynamic
equilibrium processes.
 The second law of thermodynamics is taken
 to be expressed by the Clausius
inequality\footnote{``There 
 have been nearly as many formulations of the second law 
as there have been discussions of it.'' P.W. Bridgman},  see e.g.
Rivlin~\cite{Riv70}, which states that the change in 
entropy, $\Delta S$, between times 
$t_1$ and $t_2$  in a body with volume $V$  is greater
than or equal to the time integral of the rate of heat input into the
material divided by the absolute temperature.
Hence,  in terms of the dissipation rate $\dot{\xi}$ which is the heat
  output from the material
\begin{equation} 
\Delta S \ge \int_{t_1}^{t_2} \left [ \int_V
  \frac{-\dot{\xi}}{\Theta} \ dV \right ] dt
\label{eqc}
\end{equation}
where $\Theta$ is the absolute temperature. 

From conservation of energy, Eq.~(\ref{eq11az}), 
\begin{equation} 
\Delta S \ge - \int_{t_1}^{t_2} \int_V \left [ \frac{ \dot{w} -
    \dot{\phi}  }{\Theta} 
\right ]dV  dt 
\label{eq4}
\end{equation}

The entropy change per unit volume, $\Delta {s}$, is
defined at each point of the body by 
\begin{equation}
\Delta S = \int_V    \Delta {s}   dV
\label{eq14x}
\end{equation}
so that if Eq.~(\ref{eq4}) holds for every sub-volume of the body,
then 
\begin{equation}
\Delta s \ge  - \int_{t_1}^{t_2} \left [
  \frac{\dot{w} - \dot{\phi}}{\Theta}   
\right ] dt
\label{eq15y}
\end{equation}
holds at each point $x_k$ in the body.
Note that satisfaction of  Eq.~(\ref{eq15y}) at
all points of the body  is
a more restrictive condition than satisfaction of Eq.~(\ref{eq4}). 

When Eq.~(\ref{eq15y}) holds for any time interval
$t_2 \le t \le t_1$,
\begin{equation}
 \dot w - \dot \phi + \dot{s} \Theta \ge 0 
\label{eq17h}
\end{equation}
which is the  Clausius-Duhem inequality.

The Coleman- Noll \cite{CN64} postulate is that Eq.~(\ref{eq17h})
holds for  all points in a body for all times for any 
possible loading history. This
puts the Clausius-Duhem inequality on the same
footing as the basic conservation laws of continuum mechanics even
though  the  Clausius-Duhem inequality only emerges as a statistical
limit of a large number of 
discrete events for a long time,  see e.g. Evans and Searle
\cite{Evans02}, Jarzynski \cite{Jarz10}. 

In addition to, or instead of, thermal entropy, 
configurational entropy (and configurational temperature) can be included
in a thermodynamic formulation, see for example  Bouchbinder and
Langer \cite{Bouch09},   Langer et
al. \cite{Langer10},  Falk and Langer \cite{Falk11}, McDowell
\cite{McDowell2, McDowell3}. Indeed,   it is
possible that  configurational entropy plays a
larger role than thermal entropy in 
some plastic dissipation processes.

In a purely mechanical formulation, entropy and temperature do not
enter, and the condition analogous to the Clausius-Duhem inequality,
Eq.~(\ref{eq17h}), is 
\begin{equation}
\dot{w} - \dot{\phi}=\dot{\xi} \ge 0
 \label{eq17h3} 
\end{equation}

Eq.~(\ref{eq17h3}) follows from the Clausius-Duhem inequality,
Eq.~(\ref{eq17h}), 
if $\dot{s}\Theta=0$, i.e.  if either the entropy
rate $\dot{s}=0$ or the absolute 
temperature $\Theta=0$. However:  (i) setting $\Theta=0$ in 
Eq.~(\ref{eq17h}) is problematic because $\Theta$ is in the
denominator in Eq.~(\ref{eq15y}); and (ii) setting $\dot{s}=0$ for an
irreversible 
process involving dissipation such as the evolution of plastic
deformation is questionable from a thermodynamic
perspective.  With $\dot{s}  \Theta \ge 0$,  satisfaction of
Eq.~(\ref{eq17h3}) implies satisfaction of  Eq.~(\ref{eq17h}) and then 
satisfaction of Eq.~(\ref{eq17h3}) is a sufficient, but not necessary,
condition for satisfaction of Eq.~(\ref{eq17h}),

Integrating Eq.~(\ref{eq17h3}) over the volume of the body gives
\begin{equation}
\dot{W} - \dot{\Phi}=\dot{\cal D}  \ge 0
 \label{eq17h4}
\end{equation}
If $\dot{\xi} \ge 0$ at all points of a body then
$\dot{\cal D}  \ge 0$.  
 Eq.~(\ref{eq17h4}) is less restrictive than  Eqs.~(\ref{eq17h3})
 because  Eq.~(\ref{eq17h4}) can be satisfied with  $\dot{\xi}$ negative
 at some points and non-negative at  other points.  Also, from
 Eq.~(\ref{eq4}) 
\begin{equation}
\int_V   \frac{\dot{w} -\dot{\phi}}{\Theta} \ dV  = \frac{1}{\Theta}
\left (\dot{W} - \dot{\Phi} \right ) 
\qquad    {\rm   if \ only \  if} \ \Theta={\rm constant \ in} \ V
 \label{eq17h5}
\end{equation}
Hence, the left hand side of Eq.~(\ref{eq17h4}) follows directly from
the volume integral in Eq.~(\ref{eq4}) only if the temperature $\Theta$
is constant in $V$. 

Possible restrictions that can be imposed to preclude a negative
dissipation rate in a 
purely mechanical formulation include: (i)  
$\dot{\xi} \ge 0$ for all points of the body; (ii) $\dot{\xi} \ge 0$ for
all possible deformation histories (analogous to the Coleman-Noll
\cite{CN64} postulate); and (iii) $\dot{\cal D}  \ge 0$ for the body. 

\section{Continuum slip  crystal plasticity}

The constraint imposed by  Eq,~(\ref{eq17h3})
is illustrated for a continuum slip crystal plasticity
formulation. Plastic  
deformation is represented by slips on 
specific lattice planes in specific lattice directions. The
combination of a slip direction and a slip plane normal is termed a
slip system. The total deformation is comprised of
elastic deformation of the crystal lattice and plastic deformation
arising due to slip on the specified slip systems.  Only the plastic
deformation due to slip gives rise to dissipation. 

Following Nye \cite{Nye}, the elastic
deformation of the lattice is taken to involve both lattice strain and lattice
curvature. The stored energy per unit volume, $\phi$,  is taken to
depend on the lattice 
strain $\epsilon^e_{ij}$ and the change in lattice 
curvature $\kappa^e_{ij}$.  In principle, $\phi$  could
also depend on the magnitude of slip on each slip system. That  would
lead to additional terms in equations involving the  time rate of change of
stored energy, 
but  the plastic dissipation rate expression would not be affected.  Hence, 
$\dot{\phi}$  is given by
\begin{equation}
\dot {\phi}(\epsilon^e_{ij},\kappa^e_{ij}) = \frac{\partial
  \phi}{\partial \epsilon^e_{ij}} 
\dot {\epsilon}^e_{ij}  + \frac{\partial \phi}{\partial
  \kappa^e_{ij}} \dot {\kappa}^e_{ij} 
\label{p2}
\end{equation}
The lattice curvature components, $\kappa^e_{ij}$ are  given by the
spatial gradient of the 
lattice rotation vector so that $\kappa^e_{ij}$ is not 
symmetric, i.e. $\kappa^e_{ij} \ne \kappa^e_{ji}$, Nye \cite{Nye}. 

From Eq.~(\ref{eq11az}), the work rate per unit volume can be written
as 
\begin{equation}
\dot {w}= \dot{\phi} + \dot{\xi} =\sigma_{ji} \dot {\epsilon}^e_{ij} +M_{ji}
\dot{\kappa}^e_{ij} +\dot{\xi}
\label{p1}
\end{equation}
where $\sigma_{ji}$ is the work conjugate stress and $M_{ji}$ is the
work conjugate bending moment.  From Eqs.~(\ref{p2}) and (\ref{p1}), 
\begin{equation}
\sigma_{ji} = \frac{\partial \phi}{\partial \epsilon^e_{ij}} \  ,
\ 
M_{ji} = \frac{\partial \phi}{\partial \kappa^e_{ij}} 
\label{p1xx}
\end{equation}

In continuum slip crystal plasticity, the
deformation rate due to slip is given by 
\begin{equation}
\dot{u}^p_{ij}=\sum_\beta \dot{\gamma}^{(\beta)}
s^{(\beta)}_i m^{(\beta)}_j
\label{p0a}
\end{equation}
Here,  $\dot{\gamma}^{(\beta)}$ is
the slip rate, $s^{(\beta)}_i$ is the slip direction and
$m^{(\beta)}_j$ is the slip plane normal for slip system $\beta$.

The total  strain rate, $\dot {\epsilon}_{ij}$, is 
\begin{equation}
\dot {\epsilon}_{ij}=\dot {\epsilon}^e_{ij}+\dot {\epsilon}^p_{ij}=
\dot {\epsilon}^e_{ij}+ \frac{1}{2} \left (
\dot {u}^p_{ij}+\dot {u}^p_{ji} \right )
\label{p0}
\end{equation}
with $\dot {u}^p_{ij}$ given by Eq.~(\ref{p0a}). A plastic
constitutive relation specifies $\dot{\gamma}^{(\beta)}$ as a function
of stress state and internal variables. Whether the constitutive
relation is rate independent or rate dependent is irrelevant for the
present purpose.

From Eqs.~(\ref{p0a}) and (\ref{p0})
\begin{equation}
\sigma_{ji} \dot {\epsilon}^e_{ij} = \sigma_{ji} \dot {\epsilon}_{ij}
- \sum_\beta s^{(\beta)}_i  \sigma_{ji} m^{(\beta)}_j 
\dot{\gamma}^{(\beta)} = \sigma_{ji} \dot {\epsilon}_{ij} - \sum_\beta
\tau^{(\beta)} \dot{\gamma}^{(\beta)}  
\label{p0b}
\end{equation}
where $ \tau^{(\beta)}= s^{(\beta)}_i  \sigma_{ji} m^{(\beta)}_j $ is
the Schmid resolved shear stress on slip system $\beta$. 

Because plastic deformation is only due to slip, the dissipation rate
can be written as 
\begin{equation}
\dot {\xi}=\sum_\beta  \pi^{(\beta )}  \dot
       {\gamma}^{(\beta)} 
\label{p5x}
\end{equation}

The requirement that the dissipation rate at each point $x_k$ of
the body is non-negative, Eq.~(\ref{eq17h3}), takes the form
\begin{equation}
\dot {\xi}(x_k)=\sum_\beta  \pi^{(\beta )}  \dot
       {\gamma}^{(\beta)}  \ge 0
\label{p5a}
\end{equation}

A less restrictive requirement is
\begin{equation}
 \dot{\cal D}=\int_V \pi^{(\beta)} \dot {\gamma}^{(\beta)} dV \ge 0
 \label{p5a2}
\end{equation}
with $V$ the volume of the body.

In developing crystal plasticity constitutive relations, the
requirement  typically imposed is that $\pi^{(\beta )}
\dot{\gamma}^{(\beta)} \ge 0$ for each $\beta$. However,  crystal 
structures/slip modes may be possible where at least one $\pi^{(\beta )}
\dot{\gamma}^{(\beta)} <0$ but $\dot{\xi} \ge 0$ so that 
this requirement  could be overly restrictive for
satisfying Eq.~(\ref{p5a}). 

 Nye \cite{Nye} (Appendix D) showed that, for a solid with
 dislocations,  a closed square lattice circuit in the undeformed lattice
 has a closure failure in the deformed
 curved lattice denoted by the vector $B_i$ due to dislocation
 induced   lattice
 curvature,.  Nye \cite{Nye} also
 showed that $B_i$ is  the net 
Burgers vector of the dislocations  inside a closed 
circuit  of unit area with normal $\ell_i$  and is given by 
\begin{equation}
B_i=\alpha_{ij} \ell_j
\label{p5za}
\end{equation}

Nye's dislocation
density tensor, $\alpha_{ij}$, and the lattice curvature tensor
$\kappa^e_{ij}$  are related by, Nye~\cite{Nye}, 
\begin{equation}
\kappa^e_{ij}=\alpha_{ji}- \frac{1}{2} \alpha_{nn} \delta_{ji}
\label{p6}
\end{equation}
where $\delta_{ji} $ is the Kronecker delta.

Within  the context of small deformation continuum slip plasticity theory
\begin{equation}
\alpha_{ji} = \sum_\beta e_{ikl} {\gamma}^{(\beta)}_{,k}  
s^{(\beta)}_j m^{(\beta)}_l
\label{p7}
\end{equation}
with $e_{ikl}$ being the alternating tensor, Gurtin \cite{Gur02},
Bittencourt et al. \cite{Bit03}. 

The term $M_{ji} \dot {\kappa}^e_{ij}$ in Eq.~(\ref{p1}) can be
written as
$$
 M_{ji} \dot {\kappa}^e_{ij}=  M_{ji} \dot{\alpha}_{ji} - \frac{1}{2}
   M_{nn} \dot{\alpha}_{nn} =
$$
\begin{equation}
 \sum_\beta
\left [  e_{ikl} M_{ji}
s^{(\beta)}_j m^{(\beta)}_l - \frac{1}{2} 
e_{mkl} M_{nn}  
s^{(\beta)}_m m^{(\beta)}_l \right ] \dot {\gamma}^{(\beta)}_{,k}   = \sum_\beta \chi_k^{(\beta)}
\dot {\gamma}^{(\beta)}_{,k}  
\label{p8}
\end{equation}
where
\begin{equation}
\chi_k^{(\beta)}= M_{ji} e_{ikl} 
s^{(\beta)}_j m^{(\beta)}_l - \frac{1}{2} 
e_{mkl} M_{nn}  
s^{(\beta)}_m m^{(\beta)}_l 
\label{p8z}
\end{equation}

Using Eq.~(\ref{p8z}),   Eq.~(\ref{p1}) becomes\footnote{Note that an
equation of the form of Eq.~(\ref{p8z2}) is obtained when
$\kappa^e_{ij}$ and $\alpha_{ji}$ are  related by a general linear
expression 
i.e. $\kappa^e_{ij}= R_{ijkl} \alpha_{lk}$, but with a different
expression for $ \chi_k^{(\beta)}$ from that in Eq.~(\ref{p8z}).}
\begin{equation}
\dot{w}=\left ( \sigma_{ji} \dot{\epsilon}_{ij}^e + \sum_\beta \chi_k^{(\beta)}
\dot {\gamma}^{(\beta)}_{,k}  \right )  + \sum_\beta \pi^{(\beta)}
\dot {\gamma}^{(\beta)}  
\label{p8z2}
\end{equation}
The right hand side of Eq.~(\ref{p8z2})  is the  internal work rate per unit
volume associated with  
Gurtin's \cite{Gur02} nonlocal theory of crystal plasticity with
``energetic hardening'' 
specialized to the small deformation gradient case, see also
Bittencourt et al. \cite{Bit03}.
The terms in parenthesis in Eq.~(\ref{p8z2}) are
the elastic contribution  and the last term is the plastic dissipation
rate.  With Eq.~(\ref{p0b})  used to replace the term $\sigma_{ji}
\dot{\epsilon}_{ij}^e $ in Eq.~(\ref{p8z2}),  the work rate per unit
volume expression
on which the governing equations are based is obtained. 

 The slip rate gradients, $\dot{\gamma}^{(\beta)}_{,k}$,  do not enter
 the expression for the dissipation rate because they only enter this 
 formulation  
through  their role in inducing elastic lattice bending. i.e.  via
Eq.~(\ref{p6}). In this regard it is interesting to note 
that nonphysical responses can be obtained 
when a plasticity theory includes  plastic strain rate gradient 
terms in the dissipation rate 
expression, see e.g. Fleck et al. \cite{Fleck15}.

\section{Discrete defect plasticity}

Plastic deformation of solids is almost always, if not always,
associated with the evolution of discrete defects. Plastic deformation
can then be modeled by calculating how such a 
collection of  discrete defects evolves.   Individual discrete defect
events  can take 
place over a small distance  for a short 
time (a key issue, of course, is what  constitutes ``small'' and
what constitutes ``short'').  There is no
fundamental reason to expect the Clausius-Duhem inequality,
Eq.~(\ref{eq17h}), to be satisfied or for the dissipation rate to be
non-negative for each discrete event. Indeed, enforcing the
requirement of a non-negative 
dissipation rate for each discrete defect can pose such a strong limit
on the kinetic equations describing the evolution of a
defect so as to exclude the representation of 
physically possible behaviors.

\subsection{Discrete dislocation plasticity}

The main mechanism of plastic deformation in crystalline metals at
room temperature is the glide of dislocations on specific lattice
planes. In discrete dislocation plasticity, the dislocations are
represented as line defects in a linear elastic solid. The
dislocations have a characteristic vector with the dimension of
length, ${\bf b}$,  termed the
Burgers vector. Plastic
dissipation is mainly associated with the change in configurational
energy as the dislocations glide. The dissipation rate can be written as
\begin{equation}
\dot{\cal{D}}=\sum_{K=1}^N \int_{L^{K}} P^{K} v^{K} dl
\label{eqd1a}
\end{equation}
where $N$ is the number of dislocations, $v^K$ is the dislocation
velocity, the integration is over the 
dislocation line  and $P^K$, the Peach-Koehler
(configurational) force, is given by 
\begin{equation}
P^{K}
=  \left ( \hat{\sigma}_{nb} +  \sum_{M=1,\ne K}^N
\sigma^{M}_{nb} \right ) \vert \mathbf{b}^K \vert
\label{eqd2a}
\end{equation}
Here,  ${\bf b}^K$ is the Burgers vector associated with dislocation
$K$, $ \hat{\sigma}_{nb} $ is a stress induced by the boundary
conditions, $\sigma_{nb} $ is the Schmid resolved shear stress given by
$\sigma_{nb} = \bar{b}^{K}_i \sigma_{ij}  \, n^{K}_j$ with 
$\bar{b}^{K}_i$ a unit vector in the Burgers vector direction of
dislocation $K$ and
$\vert \mathbf{b}^K \vert$ the Burgers vector amplitude. 

A kinetic relation for $v^K$ needs to be specified. For modeling 
crystals such as fcc metal crystals, a linear relation of the
form 
\begin{equation}
v^{K}=\frac{1}{C^{K}} P^{K} 
\label{eqd3aa}
\end{equation}
is typically used  where $C^K$ is the dislocation
  mobility. Eq.~(\ref{eqd3aa}) implies that plastic flow in 
such a crystal only depends on the shear stress in the slip plane
in the Burgers vector direction,  i.e. the Schmid resolved shear stress.
However, there are circumstances 
where there is evidence from atomistic calculations and experiment
that other shear stress components can play a role;  for
example,  in modeling dislocation glide when cross-slip occurs,
e.g. Hussein et al. \cite{Huss15}, Malka-Markovitz et
al. \cite{Malka21}, and  in 
modeling bcc crystals as well as 
crystals that have a  complex dislocation core structure so that 
shear stress components other than $\sigma_{nb}$ can affect
dislocation mobility,  e.g. Vitek and Paidar \cite{Vitek08}.
 In
such cases a possible kinetic relation  that has non-Schmid shear stress
dependence is of the form
\begin{equation}
v^{K}=\frac{1}{C^{K}} Q^{K} =
 \frac{1}{C^{K}} \left [ P^K + \rho^K \left
  (  \hat{\sigma}_{mt}  +  
\sum_{M=1,\ne K}^N c_M \sigma^{M}_{mt} \right ) \right ]
\label{eqd3a}
\end{equation}
where $\rho_K$ and $c_M$ are constitutive parameters, $\sigma_{mt} =
m^{K}_i \sigma_{ij}  \, t^{K}_j$with $m^{K}_i$ 
and $t^{K}_j$ unit vectors in directions  that differ from $n_i$
and/or  $\bar{b}_i$.   

The dissipation rate is 
\begin{equation}
\dot{\cal{D}}=\sum_{K=1}^N \int_{L^{K}} \left ( \frac{1}{C^{K}}
\right ) P^{K} Q^{K}  dl =\sum_{K=1}^N \dot{\cal{D}}^K
\label{eqd1c}
\end{equation}
with $C^{K}\ge 0$ and $Q^{K} =P^{K}$, $\dot{\cal{D}} $ in
Eq.~(\ref{eqd1a}) is guaranteed to be non-negative. However, with
$C^{K}\ge 0$ and 
$Q^{K} \ne P^{K}$ there is in general no guarantee that
$\dot{\cal{D}}^K\ge 0$.    
 
In this
context, the requirement analogous to the Coleman-Noll \cite{CN64}
postulate is 
\begin{equation}
\dot{\cal{D}}^K=\int_{L^{K}} \left ( \frac{1}{C^{K}}
\right ) P^{K} Q^{K}  dl  \ge 0 
\label{eqd1ca}
\end{equation}
for each $K$ and all possible stress states.

Discrete dislocation plasticity formulations with  non-Schmid
kinetic relations have been developed, e.g. by Wang and Beyerlein \cite{Wang11}
and Chaussidon et al. \cite{Chauss08}.  Wang and Beyerlein \cite{Wang11}
note that incorporating  
non-Schmid effects into the discrete dislocation formulation  is
critical for obtaining predictions  consistent with experimental
observations.  The dissipation rate associated with such formulations
has not been calculated so it is not known whether or not
Eq.~(\ref{eqd1ca}) is satisfied for the relations that have been
proposed.  In any case, for such dislocation processes it is possible to 
formulate a kinetic relation where the value of  $\dot{\cal{D}}^K$ is
negative for some $K$ for a short  time but non-negative
otherwise. 

In discrete dislocation plasticity, the most direct calculation is for
$\dot{\cal{D}}^K$. However, $\dot{\xi}$ can also be calculated as in
Benzerga et al. \cite{Amine03}, but that is a more complex calculation. If
$\dot{\cal{D}}^K<0$ then necessarily $\dot{\xi}<0$ at one or more 
points of the body.

\subsection{Discrete shear transformation plasticity}

Plastic deformation of amorphous materials, such as metallic glasses,
typically occurs 
via local atomic re-arrangements termed shear transformation
zones (STZs).  In the context of continuum mechanics, STZs have been 
modeled as transforming Eshelby \cite{Eshelby57,Eshelby59}
inclusions. A transforming Eshelby 
\cite{Eshelby57,Eshelby59} inclusion is an
elliptical sub-volume of  a uniform linear elastic solid that undergoes a
transformation such that if the inclusion were removed from the
matrix, it would undergo a uniform strain $\epsilon^*_{ij}$  with zero
stress.   However, due to the constraint
of the surrounding matrix material both the material inside the
inclusion and the material outside the inclusion are stressed. 

The dissipation rate $\dot{\cal D}$ is calculated via Eq.~(\ref{eq0})
as, Vasoya et al. \cite{Vasoya20}, 
\begin{equation}
\dot{\cal D} = \sum_{K=1}^N Z^K_{ij}
\dot{\epsilon}^{*K}_{ij}
\label{eqzza}
\end{equation}

For a body with a collection of $N$ STZs  and with a stress field
$\sigma^0_{ij}(x_k,t)$ arising from the imposed loading 
\begin{equation}
Z^K_{ij}=\int_{V^{K}} \left ( \sigma^0_{ij}(x_k,t)+\sum_{J=1}^N
\sigma^{T_J}_{ij} (x_k,t) \right ) dV
\label{eqzz}
\end{equation}
where the integration is over the volume of STZ $K$ and
$\sigma^{T_J}_{ij}(x_k,t)$ is the transformation stress associated
with STZ $J$ that is uniform in STZ $J$ for an elliptical inclusion,
Eshelby \cite{Eshelby57,Eshelby59}.

A non-negative dissipation rate $\dot{\cal D}$ requires
\begin{equation}
\dot{\cal D}=\sum_{K=1}^N Z^K_{ij} \dot{\epsilon}^{*K}_{ij} \ge 0
\label{pd1a}
\end{equation}

The requirement analogous to the Coleman-Noll \cite{CN64} postulate is
\begin{equation}
Z^K_{ij} \dot{\epsilon}^{*K}_{ij} \geq 0 \quad {\rm for \ each \ K}
\label{pd1aa}
\end{equation}
and the condition Eq.~(\ref{pd1aa}) is satisfied if
\begin{equation}
Z^K_{ij} = K^K_{ijkl} \dot{\epsilon}^{*K}_{kl} 
\label{pd1ab}
\end{equation}
where each $K^K_{ijkl}$ is a positive definite fourth order tensor so
that
\begin{equation}
\dot{\cal D}=\sum_{K=1}^N \dot{\epsilon}^{*K}_{ij}  K^K_{ijkl}
\dot{\epsilon}^{*K}_{kl} 
\label{pd1ab2}
\end{equation}

Employing Eq.~(\ref{pd1ab}) for a shear transformation or a
volumetric transformation limits the maximum allowable transformation
strain to values that are smaller than indicated by atomistic
calculations and experiment, as discussed by Vasoya et
  al. \cite{Vasoya20}.  A kinetic equation for an Eshelby
  \cite{Eshelby57,Eshelby59} inclusion 
  undergoing a plane strain shear transformation that allows for a
  negative dissipation rate for a short time  is given by
  Needleman \cite{EMC23}, Eq. (64), that, in principle, allows
  for larger transformation strains to be achieved. However, whether
  such a relation is physically appropriate and what conditions need
  to be placed on the kinetic equation parameters to ensure stability
  are not known. 

As for discrete dislocation plasticity, it is simpler to calculate
$\dot{\cal D}^K$ than it is to calculate $\dot{\xi}$ at each point of
the body.

\section{Plasticity and a Negative Dissipation Rate}

At least in some contexts, the requirement, analogous to the
Coleman-Noll \cite{CN64} postulate, that the dissipation rate is
non-negative for all possible processes is overly
restrictive. Presuming that requiring a non-negative dissipation rate 
is, in some sense, a stability requirement, a challenge is to
develop a dissipation rate restriction that allows for a negative
dissipation rate 
but that guarantees stability. This is of particular significance for
discrete defect plasticity.

\begin{figure*}[htb!]
\begin{center}
\subfigure[]
{\resizebox*{73mm}{!}{\includegraphics{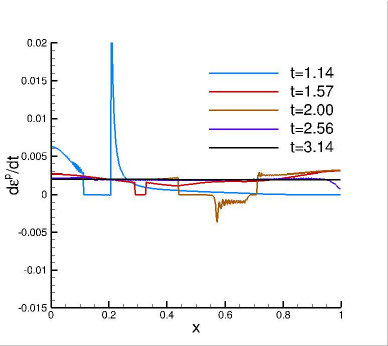}}}
\subfigure[]
{\resizebox*{73mm}{!}{\includegraphics{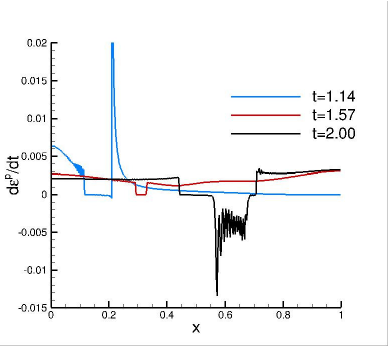}}}
\end{center}
\caption{ Plastic strain rate as a function of position at various
  normalized times for a one dimensional wave  calculation. (a)
  Stability with a
  negative dissipation rate  over a small region  for a short  time. (b)
  Instability with a larger negative dissipation rate.  From Needleman
  \cite{EMC23}.} 
\label{1da}
\end{figure*}

Although conditions to impose for maintaining 
stability in plasticity calculations where a negative
dissipation rate occurs are not available,   a simple
one dimensional wave propagation calculation showed that it is
possible for stability to be be maintained with a negative 
dissipation rate occurring for a short time, Needleman \cite{EMC23}.
Fig.~\ref{1da}(a) shows an instability developing at $t=2.0$
around $x=0.6$ but that instability is damped out at a later times. An
instability does develop around $x=0.6$ in a calculation
with a more negative 
dissipation rate, Fig.~\ref{1da}(b), so that there is a limit to a
negative dissipation 
rate that can occur with stability maintained. 

There is anecdotal evidence (i.e. seen in some calculations that I was
involved in) that a negative dissipation
rate can occur in a discrete defect plasticity calculation with no
indications of an instability. The dissipation rate restriction 
required for 
stability to be maintained while one or more discrete events have a
negative dissipation rate has not yet been developed.   

In this regard, it should be mentioned, that there is the possibility
that a full thermodynamic discrete defect plasticity framework,
including  temperature and 
entropy,  is needed to provide a formulation that appropriately accounts
for a negative dissipation rate.
In such a framework, the Clausius-Duhem inequality,
Eq.~(\ref{eq17h}), can be satisfied even when the dissipation rate is
negative. Whether or not a formulation that satisfies the
Clausius-Duhem inequality 
but that gives  a negative dissipation rate can maintain stability
remains to be determined. 

One use of discrete defect plasticity modeling is to develop
the background for a higher level crystal plasticity constitutive
relations and the consequence of allowing the possibility of a
negative dissipation rate for individual discrete events
for the higher scale modeling remains to be investigated. Quite
generally there is a need for continuum mechanics formulations that
maintain stability and, consistent with statistical mechanics, allow for a
negative dissipation rate in  a 
small region for a short time.

\section{Concluding remarks}

\begin{description}

\item{1.}   The Clausius-Duhem inequality emerges from a statistical
description of interacting discrete entities in the limit of a large
number of  discrete entities interacting  for a long time. It is not a
fundamental physical requirement for all discrete events.

\begin{description}

\item{(i).} In plasticity formulations, a restriction on dissipation
  rate can be phrased in terms 
  of: (i) the   pointwise dissipation rate $\dot{\xi}$; or (ii) the volume
  integrated dissipation rate $\dot{\cal D}$.   The requirement that
  $\dot{\xi}$ or $\dot{\cal D}$ be non-negative for all possible
  deformation histories can be overly restrictive in some contexts.
  In such a context, the appropriate dissipation rate measure 
  to restrict and the appropriate restriction to impose remain to
  be determined for  various plasticity formulations.

\end{description}

\item{2.} The requirement in a purely mechanical  theory of plasticity that the
plastic dissipation rate is non-negative at each point of the body 
is related to, but not 
strictly derived from, the 
Clausius-Duhem inequality except in the special, and questionable,
case of a plastic dissipative process with no change in entropy.

\item{3.} When plastic deformation is characterized by a collection of
  discrete events, the requirement that each discrete event gives rise to
  a non-negative dissipation rate can be overly restrictive.

\begin{description}

\item{(i)} In particular, for discrete defect plasticity, the requirement of a
non-negative dissipation rate for each discrete event imposes strong
restrictions on discrete 
defect evolution equations that can preclude modeling experimentally
observed and/or atomistically predicted processes.

\end{description}

\item{4.}  When  a negative dissipation rate  is permitted
  for a small region for a short time, the requirement for
  ensuring stability remains to be developed.

\end{description}

\section*{Acknowledgment}
I am  pleased to thank  Professor Ahmed Benallal of LMPS, Universit\'e
Paris-Saclay, Centrale Supelec, ENS Paris-Saclay, CNRS 
for helpful comments
and suggestions on a previous draft.

%\section*{References}

\end{document}